\documentclass[twocolumn,epj]{svjour}          
\pdfoutput=1 

\usepackage{preprintcover}  
\PreprintCoverPaperTitle{Search for pair-produced massive coloured scalars in four-jet final states with the ATLAS detector in proton-proton collisions at $\sqrt{s} = 7$\,\TeV}  
\PreprintIdNumber{CERN-PH-EP-2012-256}  
\PreprintCoverAbstract{A search for pair-produced massive coloured scalar particles decaying to a four-jet final state is performed by the ATLAS experiment at the LHC in proton-proton collisions at $\sqrt{s} = 7$\,\TeV.  
The analysed data sample corresponds to an integrated luminosity of 4.6\,\ifb.
No deviation from the Standard Model is observed in the invariant mass spectrum of the two-jet pairs. 
A limit on the scalar gluon pair production cross-section of 70\,pb (10\,pb) is obtained at the 95\% confidence level for a scalar gluon mass of 150\,\GeV\ (350\,\GeV).
Interpreting these results as mass limits on scalar gluons, masses ranging from 150\,\GeV\ to 287\,\GeV\ are excluded at the 95\% confidence level.}  
\PreprintJournalName{EPJC Letters}  

\usepackage{url}

\usepackage{graphicx}
\graphicspath{{figures/}}
\DeclareGraphicsExtensions{.pdf}
\usepackage{subfigure}
\usepackage{mathrsfs}
\usepackage{authblk}
\usepackage{textcomp}
\usepackage[switch]{lineno} 

\usepackage{graphicx}
\usepackage{textpos}
\usepackage{amssymb}
\usepackage{dcolumn}
\usepackage{bm}
\usepackage{multirow}
\usepackage{epsfig}
\usepackage{subfigure}
\usepackage{cite}

\def\MeV{\ifmmode {\mathrm{\ Me\kern -0.1em V}}\else
                   \textrm{Me\kern -0.1em V}\fi}%
\def\GeV{\ifmmode {\mathrm{\ Ge\kern -0.1em V}}\else
                   \textrm{Ge\kern -0.1em V}\fi}%
\def\TeV{\ifmmode {\mathrm{\ Te\kern -0.1em V}}\else
                   \textrm{Te\kern -0.1em V}\fi}%
\def\ifb{\mbox{fb$^{-1}$}}
\def\ipb{\mbox{pb$^{-1}$}}
\def\pt{\ensuremath{p_{\mathrm{T}}}} 
\def\Alpgen{{\sc Alpgen}}
\def\Pythia{{\sc Pythia}}
\def\Herwig{{\sc Herwig}}
\def\Jimmy{{\sc Jimmy}}

\begin{document}

\title{Search for pair-produced massive coloured scalars in four-jet final states with the ATLAS detector in proton-proton collisions at $\sqrt{s} = 7$\,\TeV}
\titlerunning{Search for massive coloured scalars with the ATLAS detector}
\author{The ATLAS Collaboration}
\institute{}
\date{}

\abstract{
A search for pair-produced massive coloured scalar particles decaying to a four-jet final state is performed by the ATLAS experiment at the LHC in proton-proton collisions at $\sqrt s = 7$\,\TeV.  
The analysed data sample corresponds to an integrated luminosity of 4.6\,\ifb.
No deviation from the Standard Model is observed in the invariant mass spectrum of the two-jet pairs. 
A limit on the scalar gluon pair production cross-section of 70\,pb (10\,pb) is obtained at the 95\% confidence level for a scalar gluon mass of 150\,\GeV\ (350\,\GeV).
Interpreting these results as mass limits on scalar gluons, masses ranging from 150\,\GeV\ to 287\,\GeV\ are excluded at the 95\% confidence level.
\PACS{ {12.60.Jv}{Supersymmetric models}, \and {14.80.Ly}{Supersymmetric partners of known particles} }
}

\maketitle

Massive coloured scalar particles that decay into gluons are predicted 
in several extensions of the Standard Model (SM). The most prominent examples are
the scalar partners of a Dirac gluino called scalar gluons (`sgluons') in extended supersymmetric models like the 
$\mathcal{N}$=1\,/\,$\mathcal{N}$=2 hybrid model~\cite{Fayet:1975yi,Choi:2008ub} or the R-symmetric MSSM~\cite{Kribs:2007ac,Plehn:2008ae}.
These particles are also present in compositeness models~\cite{Kilic:2008pm, Kilic:2008ub, Schumann:2011ji, Kilic:2009mi, Bai:2011mr} where they are called hyperpions.
While single production of sgluons is possible, the production cross-section depends strongly on the masses of the supersymmetric particles and, in typical supersymmetric scenarios, is of the same order as the pair production cross-section. 
On the contrary the pair production cross-section does not, at leading order, depend on supersymmetric parameters except for the sgluon mass. 
Since the sgluon has positive R-parity~\cite{Farrar:1978xj} and since the sgluon coupling to quark-antiquark pairs is suppressed by the quark mass, light sgluons, 
i.e. sgluons with masses of the order of 100\,\GeV, are expected to decay to two gluons with a branching ratio close to one~\cite{Choi:2008ub,Plehn:2008ae}.
Pair production of sgluons each decaying to two gluons, leading to a four-jet final state, is therefore used as a benchmark process.
ATLAS has previously searched for signatures of these particles in the dataset of 34\,\ipb\ recorded in 2010~\cite{Aad:2011yh},  excluding at the 95\% confidence level (CL) sgluons with masses of 100\,\GeV\ to 180\,\GeV, with the exception of a mass window of 5\,\GeV\ around 140\,\GeV.
The search described in this paper, using data recorded in 2011, explores the mass region from 150\,\GeV\ up to about 300\,\GeV.

The strategy of the analysis is to first reconstruct the two sgluon candidates. The relative mass difference and the $\Delta R$
between the two jets associated to a reconstructed sgluon are used to select well reconstructed sgluons with a small mass difference,
where $\Delta R=\sqrt{(\Delta \phi)^2+(\Delta \eta)^2} $ with $\Delta \phi$ and $\Delta \eta$ being the difference in azimuth and pseudorapidity of the two jets.
The distribution of the reconstructed average mass of the two sgluon candidates is then analysed for evidence of a signal 
with a fit to the background plus a signal template of variable strength. 
The background for this search is the SM multijet background, due to its large cross section. 

ATLAS is a multipurpose detector~\cite{Aad:2008zzm,Aad:2009wy} with nearly $4\pi$ coverage in solid angle.
The inner detector, consisting of silicon pixel and microstrip detectors as well as a transition radiation tracker,
is immersed in a 2~T axial magnetic field.
In the pseudorapidity\footnote{ATLAS uses a right-handed coordinate system with its origin at the nominal interaction point (IP)
in the centre of the detector and the $z$-axis along the beam pipe. The $x$-axis points from the IP to the centre of
the LHC ring, and the $y$-axis points upward. Cylindrical coordinates $(r,\phi)$ are used in the transverse $(x$-$y)$ plane,
$\phi$ being the azimuthal angle around the beam axis. The pseudorapidity is defined in terms of the polar angle
$\theta$ as $\eta=-\ln\tan(\theta/2)$.} region $|\eta|<3.2$, high-granularity lead/liquid-argon (LAr)
electromagnetic (EM) sampling calorimeters are used. An iron/scintillator-tile calorimeter
provides hadronic coverage for $|\eta|<1.7$. The end-cap and forward regions, spanning
$1.5<|\eta|<4.9$, are instrumented with LAr calorimeters for both EM and hadronic measurements.
The calorimeters are surrounded by a muon spectrometer which consists of three large superconducting toroids,
a system of precision tracking chambers, and fast detectors for triggering.
ATLAS uses a three-level trigger system. The first-level trigger is implemented in custom hardware, 
the other two trigger levels are implemented in software running on commercially available PC farms. 

\begin{sloppypar} 
A data-driven method is used for the background estimation. 
The method is developed and validated on SM multijet Monte Carlo (MC) samples.
The samples are also used to determine a systematic error on the background determination. 
To incorporate detector effects, these events are passed through a detailed simulation of the ATLAS detector~\cite{Aad:2010ah} based on {\sc Geant4}~\cite{Agostinelli:2002hh}.
\Alpgen~\cite{Mangano:2002ea} SM multijet samples are generated with the MLM matching scheme~\cite{Hoche:2006ph} and interfaced to 
\Herwig~\cite{Corcella:2002jc} for the parton shower and fragmentation processes and to 
\Jimmy~\cite{Butterworth:1996zw} for the simulation of the underlying event. The \Alpgen\ samples are generated with the CTEQ6L1 parton distribution functions (PDFs)~\cite{Lai:2010vv} (underlying event tune AUET2-CTEQ6L1~\cite{ATL-PHYS-PUB-2011-009}).
\mbox{\Pythia~\cite{Sjostrand:2006za}} SM multijet samples are generated with the \mbox{LO$^*$~MRST PDFs~\cite{Sherstnev:2007nd}} (underlying event tune AUET2B LO$^{**}$~\cite{ATL-PHYS-PUB-2011-009}).
The sgluon pair production differential cross-section~\cite{Choi:2008ub} is implemented as an external process interfaced to \Pythia.
The decay of the sgluons as well as fragmentation and hadronization are performed by \Pythia. For the normalisation of the signal a next-to-leading-order (NLO) cross-section is used~\cite{GoncalvesNetto:2012nt}.
\end{sloppypar} 

Signal samples of 40k to 120k events each were generated with sgluon masses of 150, 200, 250, 300 and 350\,\GeV\ and passed through the fast simulation of the ATLAS detector. Additionally the fast simulation was verified by passing some samples through the
detailed simulation which leads to results with good agreement.

\begin{figure}[htbp]
	\center
	\includegraphics[width=\columnwidth]{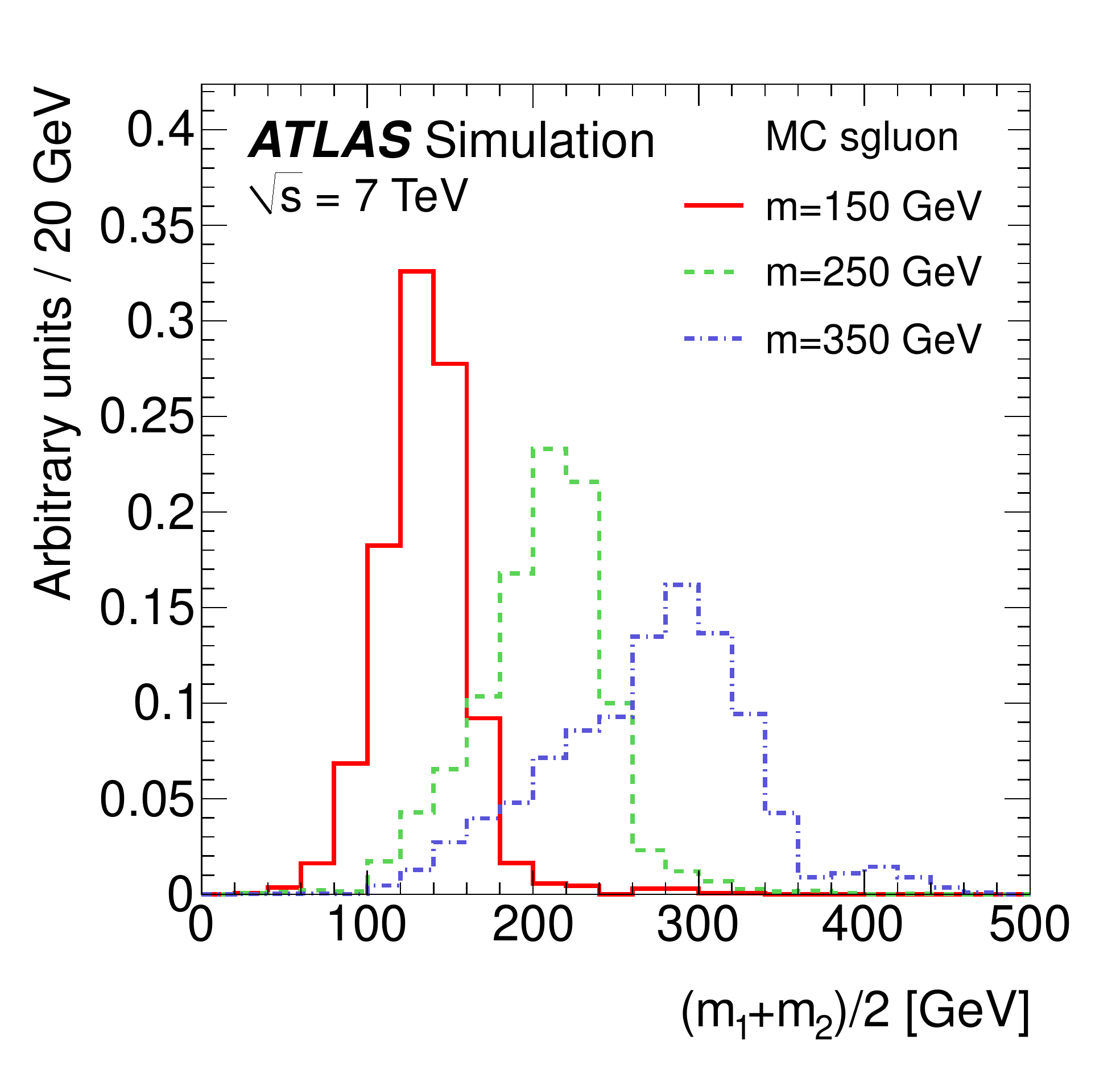}
	\caption{The reconstructed average mass distribution after all analysis cuts for the signal samples with $m_{\mathrm{sgluon}}$=150, 250 and 350\,\GeV. The curves are normalized to unit area.}
	\label{fig:fitsig}
\end{figure}

Jets are reconstructed using the anti-$k_t$ jet clustering algorithm~\cite{Cacciari:2008gp} with a radius parameter of 0.4. 
The inputs to the jet algorithm are three-dimensional clusters~\cite{Lampl:2008zz} formed from energy deposits in the calorimeters. 
The jets are calibrated using transverse momentum ($p_\mathrm{T}$) and $\mathrm{\eta}$-dependent correction factors based on MC simulations 
and validated by test beam and collision data studies~\cite{Aad:2012vm}.  
Quality criteria are applied to reject jets produced by non-collision backgrounds~\cite{ATLAS-CONF-2012-020}.
Such jets are typically produced by hardware problems in the calorimeters, LHC beam-gas interactions or cosmic-ray induced showers.
The jet energy resolution for a jet with a \pt\ of 80~\GeV\ is about 11\%~\cite{Aad:2012ag}.

The analysis uses collision data collected in the year 2011 at a centre-of-mass energy of $\sqrt s = 7$\,\TeV\ and corresponding 
to an integrated luminosity of 4.6\,\ifb.
The data were recorded with a multijet trigger requiring at least four jets. Reflecting the threshold of the third level trigger of 45~\GeV, 
the trigger efficiency does not depend strongly on the \pt\ of the four highest-\pt\ jets for $\pt > 80$~\GeV.
To ensure full trigger efficiency for the analysis, the four highest-\pt\ jets in a selected event are required to be separated from each other by $\Delta R> 0.6$.
The resulting trigger efficiency of at least 99\% obtained with simulated events was verified in data with the use of a single-electron trigger.
The average number of proton-proton interactions in the same event (pile-up) has increased to about 12~events over the course of the run. 
Simulated minimum-bias events are overlayed onto the simulated samples of the hard-scattering
processes. The resulting events are reweighted to reproduce the luminosity profile of the data.
The requirement on the \pt\ of the jets makes the analysis robust with respect to the increase of pile-up.

\begin{sloppypar} 
A selected event must contain at least one reconstructed primary vertex with five or more associated tracks each having $\pt>400$\,\MeV. 
At least four jets with $\pt>80$\,\GeV\ and $|\eta|< 1.4$ are required, since the signal is produced centrally, in contrast to the SM multijet background. 
These selection criteria together with the trigger requirements are referred to as preselection in the following.
To improve the sensitivity of the analysis, the jet \pt\ threshold is defined as a function of the probed sgluon mass.
The \pt\ of the fourth highest-\pt\ jet is required to be greater than the maximum between 80\,\GeV\ and 0.3 times the sgluon mass plus 30\,\GeV\ \mbox{($\pt(4^{th}\textrm{jet})>\max(0.3\times m_{\mathrm{sgluon}} + 30, 80)\GeV$)}.
The stringent \pt\ requirements select sgluons produced with a boost and hence lead to an accumulation of jet pairs from sgluon decays with $\Delta R\approx 1$.
Taking advantage of this property to reconstruct the two sgluon candidates, the four highest-\pt\ jets in a preselected event are paired by minimising the quantity \mbox{$|\Delta R_{\mathrm{pair1}}-1|+|\Delta R_{\mathrm{pair2}}-1|$.}
The event is rejected if, for the chosen combination, a jet pair has $\Delta R>1.6$.
The reconstructed masses of the sgluons are denoted $m_1$ and $m_2$.
The reconstructed average mass is $(m_1+m_2)/2$.
The scattering angle is defined as the angle between the direction of motion of the reconstructed sgluons in the rest frame of the four highest-\pt\ jets and the boost direction between the lab frame and the rest frame of the four highest-\pt\ jets.
The magnitude of the cosine of the scattering angle $|\cos(\theta^*)|$ is required to be less than 0.5.
In fact the SM multijet background is peaked in the forward region, reflecting \textit{t}-channel gluon exchange, while the signal is produced centrally due to the (mainly) \textit{s}-channel production and scalar nature of the sgluon.
Finally, to further improve the rejection of the SM multijet background, the relative difference between the two reconstructed masses $|m_1 - m_2|/(m_1+m_2)$ is required to be less than 0.15.
The requirement on the relative mass difference selects well reconstructed 
events increasing the bulk of the signal distribution with respect to the tails.
Loosening the requirement, i.e., accepting events with larger mass differences, leads to an increase
of signal events but also an increase of background events. 
The selection that increases the number
of events in the control regions without decreasing the sensitivity is chosen.
As the requirement on the 4$^{th}$ jet is mass dependent, 
a selection efficiency of 0.6\% is achieved for all simulated samples of the different sgluon masses.
\end{sloppypar} 

The reconstructed average mass distribution after all cuts is shown for sgluon signals with varying masses in Fig.~\ref{fig:fitsig}. The natural width of the sgluon in this mass range is negligible and the width of the observable is entirely dominated
by the instrumental mass resolution. As the mass of the sgluon increases, the requirement on the transverse momentum of the jets becomes relatively less stringent. An increase of 100\,\GeV\ in the sgluon mass leads to an increase of only 30\,\GeV\ in the requirement on the jet transverse momenta. As a consequence the radiative tails to lower masses are more evident for higher masses in the figure, as they are less sculpted by the \pt\ cut.  

\begin{figure*}[htb]
   \begin{minipage}[c]{.5\linewidth}
      \includegraphics[scale=0.42]{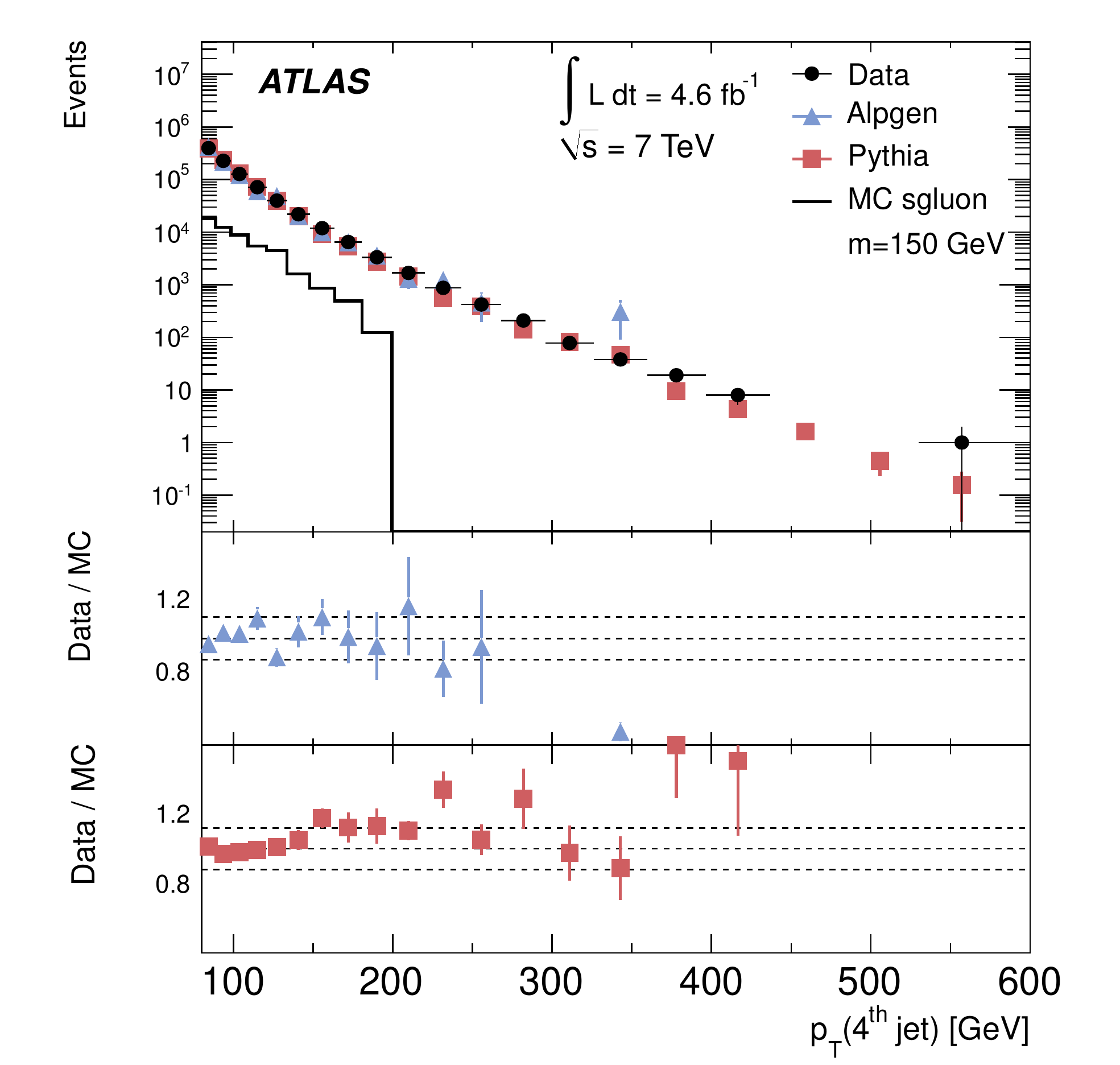}
   \end{minipage} \hfill
   \begin{minipage}[c]{.5\linewidth}
      \includegraphics[scale=0.42]{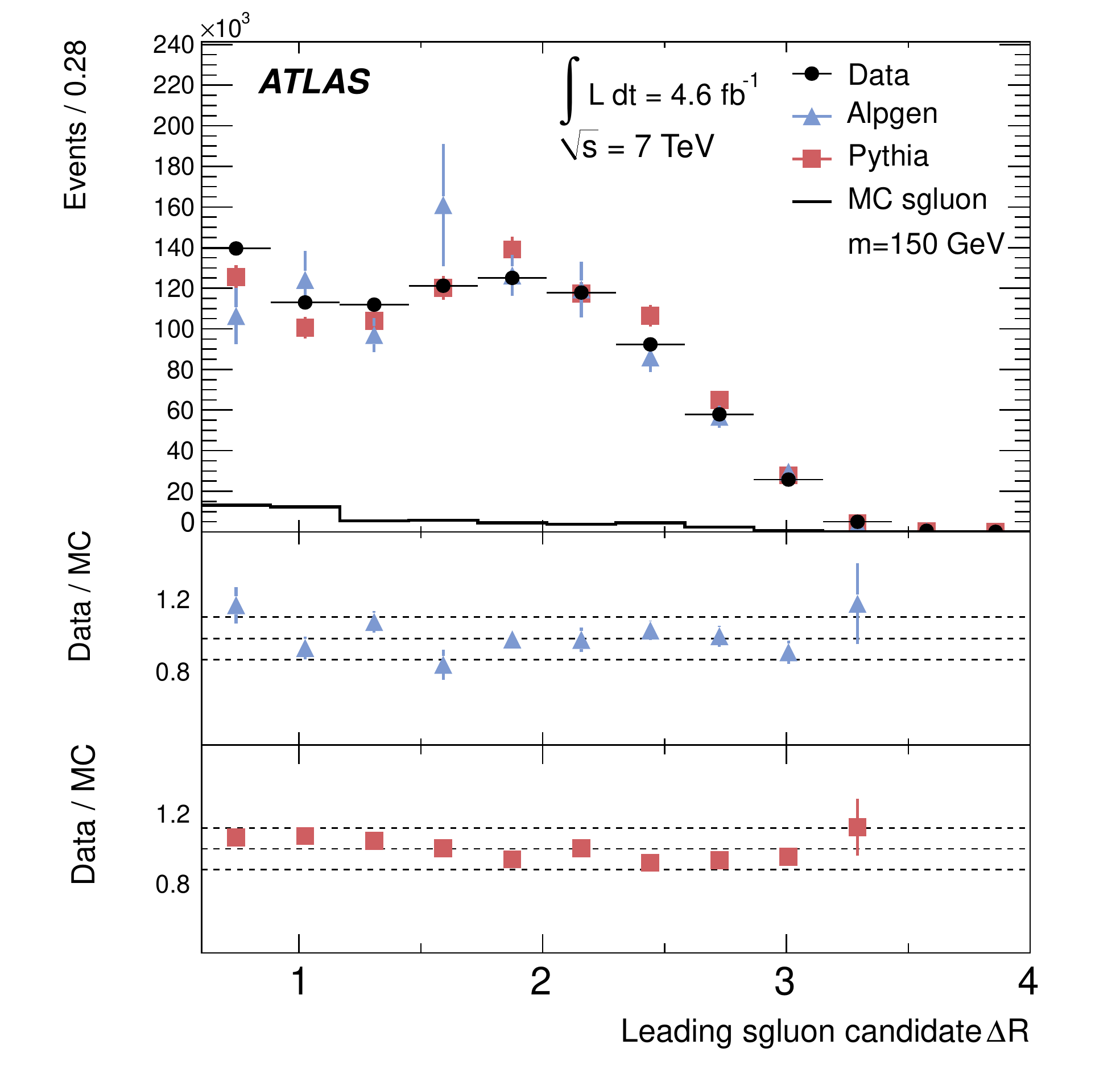}
   \end{minipage}
   \begin{minipage}[c]{\linewidth}
   \begin{picture} (12.,10.)(0.,0.)
		\setlength{\unitlength}{1.0cm}
		\put (4.5, 0.25){(a)}
		\put (13.5, 0.25){(b)}
   \end{picture}
   \end{minipage}
   \begin{minipage}[c]{.5\linewidth}
      \includegraphics[scale=0.42]{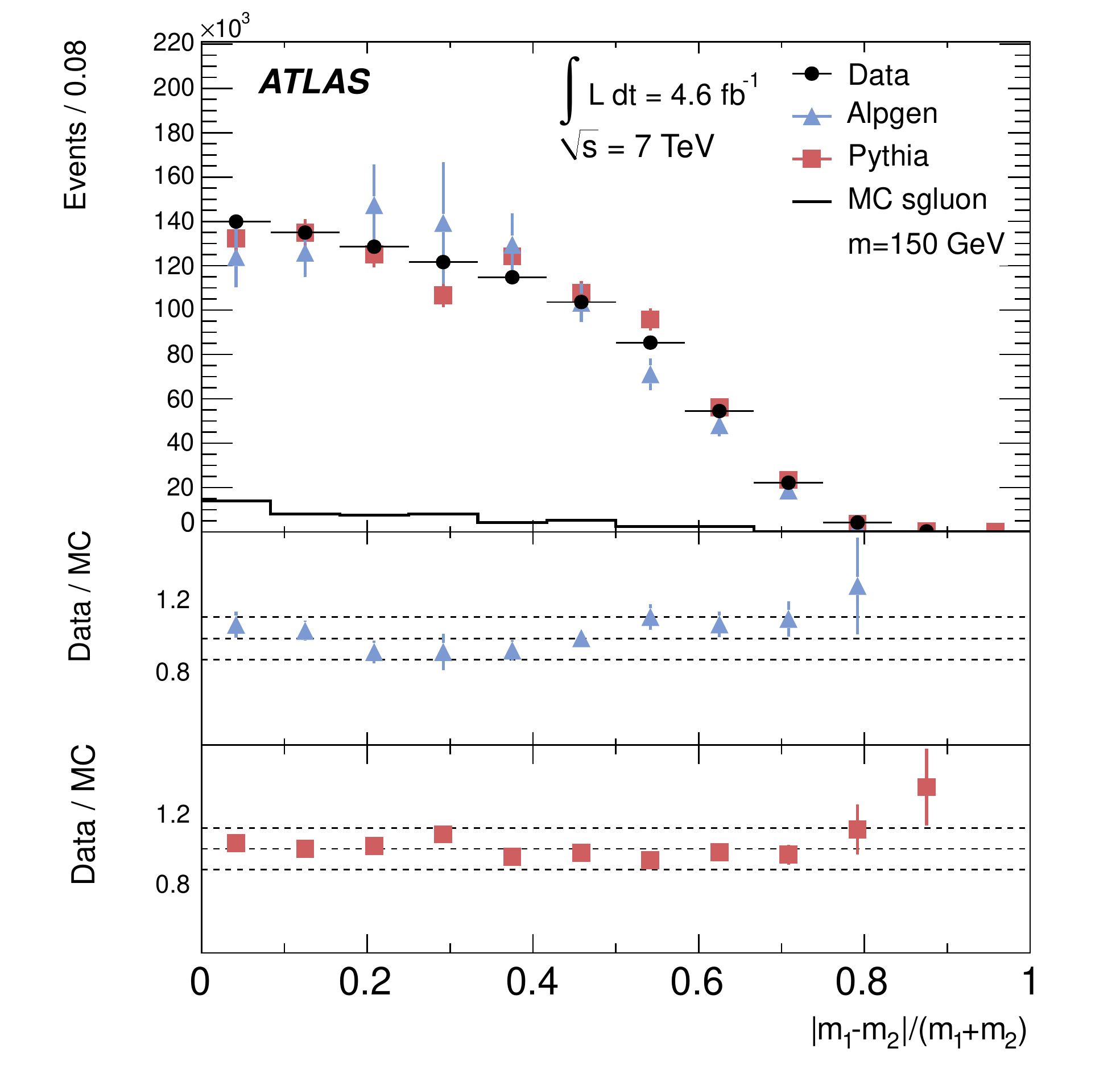}
   \end{minipage} \hfill
   \begin{minipage}[c]{.5\linewidth}
      \includegraphics[scale=0.42]{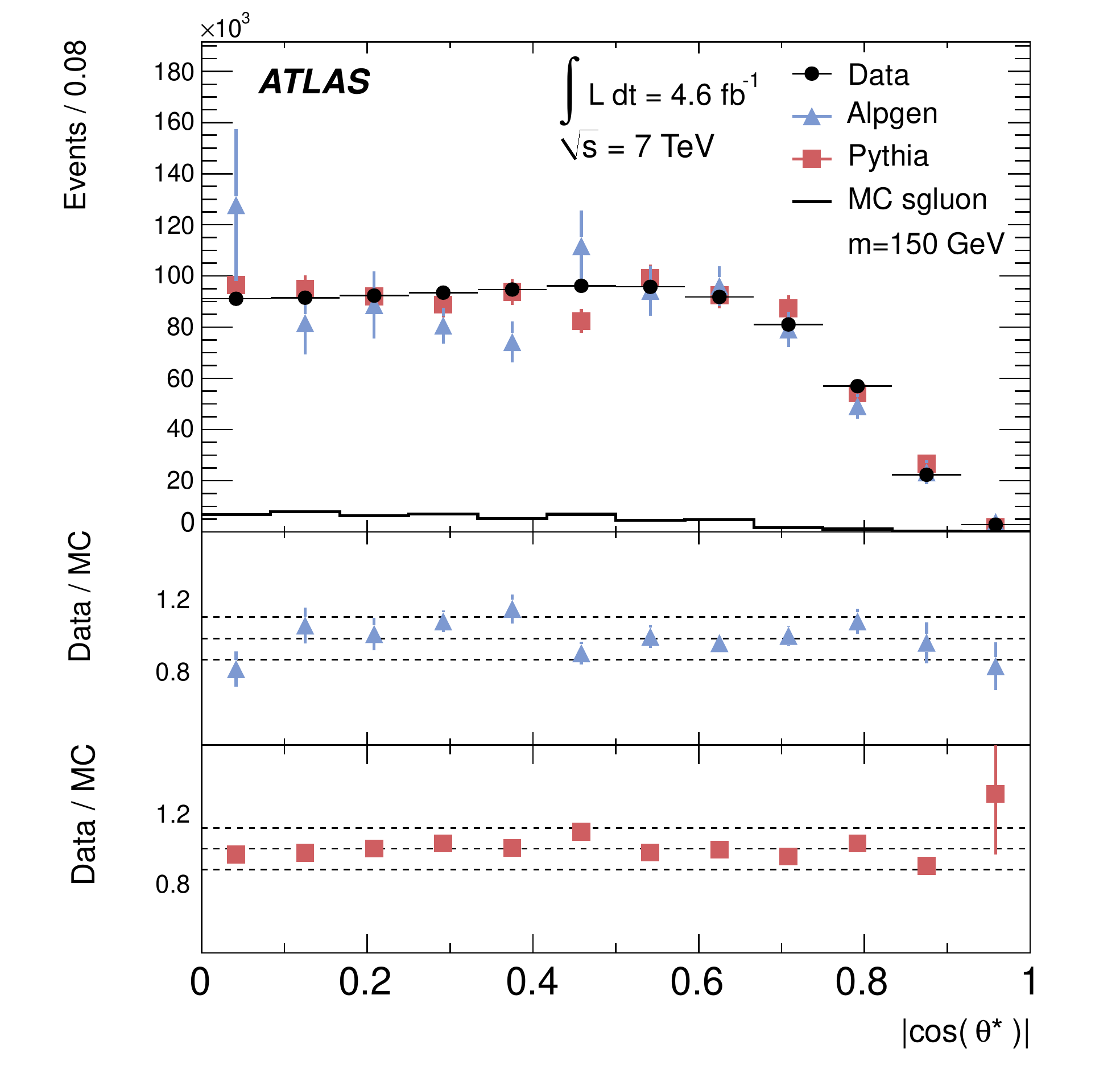}
   \end{minipage}
   \begin{minipage}[c]{\linewidth}
   \begin{picture} (12.,10.)(0.,0.)
		\setlength{\unitlength}{1.0cm}
		\put (4.5, 0.25){(c)}
		\put (13.5, 0.25){(d)}
   \end{picture}
   \end{minipage}
   \caption{The kinematic variables of the analysis are shown after applying the preselection and pairing the four highest-\pt\ jets:
(a) is the transverse momentum of the fourth highest-\pt\ jet. 
(b) is the $\Delta R$ between the two jets of the reconstructed sgluon candidate with the highest transverse momentum jet. 
(c) is the relative mass difference. 
(d) is the cosine of the scattering angle in the four-jet centre-of-mass frame.
The black histogram is the signal for a sgluon mass of 150\,\GeV\ normalised to the NLO cross-section.
Data (dots) are compared to the \Alpgen\ (triangles) and \Pythia\ (rectangles) SM multijet samples where the MC samples are normalised to the data.
The ratio data/MC is also shown separately for \Alpgen\ and \Pythia\ with its statistical uncertainty.}
   \label{fig:data_mc}
\end{figure*}
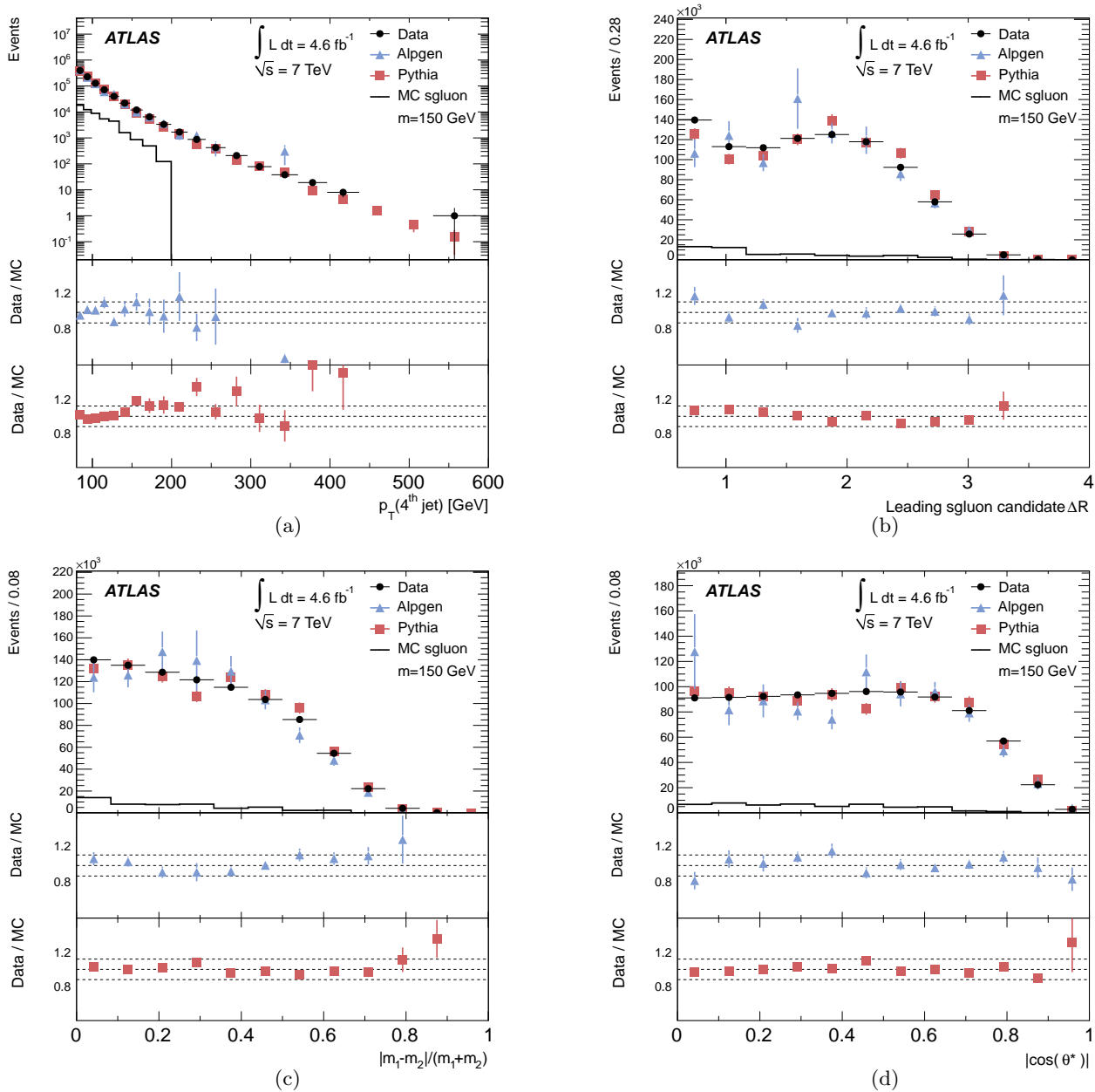

After applying preselection and jet pairing, the primary variables used in the analysis are compared to the \Alpgen\ and \Pythia\ multijet simulation. 
Backgrounds other than SM multijet events are estimated to be smaller than 1\% of the total background sample and are thus neglected. 
Since the analysis requires at least four jets, \Alpgen\ is expected to give a better description than \Pythia,
which generates the third and fourth jets via a parton shower.

The MC samples are normalised to the data after the preselection described above.
The normalisation factor of 1.25 obtained for \Alpgen\ is compatible with the one (1.26) obtained in Ref.~\cite{Aad:2011yh}.
For \Pythia\ the normalisation factor is 0.75, close to the value of 0.65~\cite{Aad:2011tqa}
obtained with the 2010 data using a different tune (AMBT1).
After this normalisation, for the transverse momentum of the 4$^{th}$ jet~(Fig.~\ref{fig:data_mc}(a)), the separation of the two jets of the sgluon candidate with the
highest transverse momentum (Fig.~\ref{fig:data_mc}(b)), the relative mass difference (Fig.~\ref{fig:data_mc}(c)) and the cosine of the scattering
angle (Fig.~\ref{fig:data_mc}(d)), the agreement between the data and the MC simulations is, in general, at the 20\% level.
To reduce the dependence on simulation, the background is estimated from data, taking advantage of the kinematic properties of the sgluon signal.
Only the systematic error on the method is taken from the Monte Carlo studies.

The main discriminating variable for the analysis is the reconstructed average mass. To determine the background normalisation as well as the shape of the background in the signal region, an ABCD method is used. The data sample is divided into one signal region (A) and three background-dominated regions (B, C and D).
The variables used to define the four regions are $|\cos(\theta^*)|$ and $|m_1 - m_2|/(m_1+m_2)$. 
The regions defined in Table~\ref{tab:defabcd} are chosen as a compromise between the statistical significance of the signal in region A and the statistical uncertainty in the regions B, C and D which feeds into the uncertainty on the background prediction. 

\begin{table}[htt]
\caption{\small Definition of the four regions for the background determination. Region A is the signal region.}
\centering
\renewcommand{\arraystretch}{1.1}
\begin{tabular}{c||c|c}
\hline
\hline
Region   &  $|\cos(\theta^*)|$ & $|m_1 - m_2|/(m_1+m_2)$ \\
\hline
A & $<0.5$ & $<0.15$ \\
B & $>0.5$ & $<0.15$ \\
C & $<0.5$ & $>0.15$ \\
D & $>0.5$ & $>0.15$ \\
\hline
\hline
\end{tabular}
\label{tab:defabcd}
\end{table}

\begin{figure*}[htb]
   \begin{minipage}[c]{.5\linewidth}
   	  \includegraphics[scale=0.42]{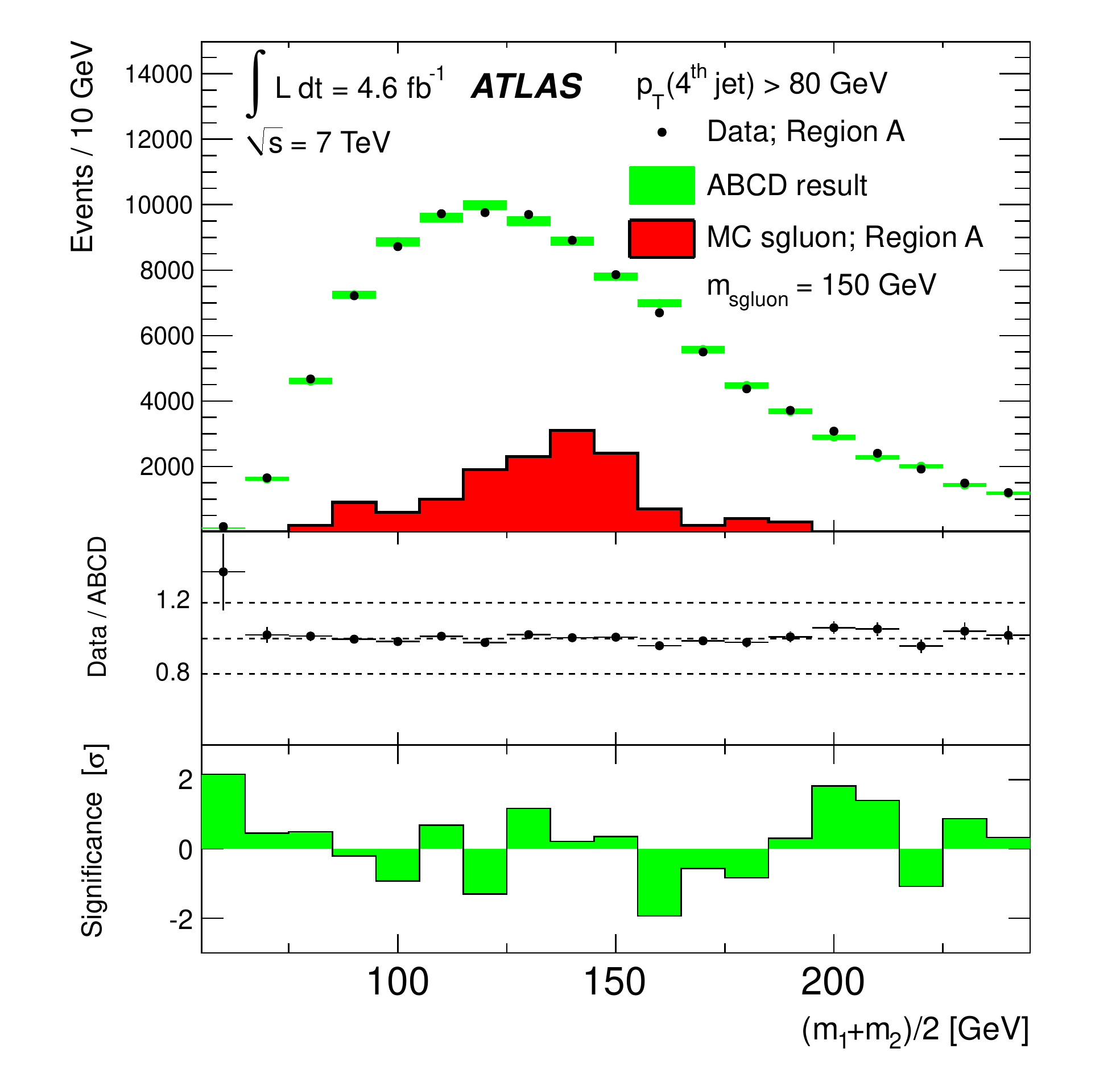}
   \end{minipage}
   \begin{minipage}[c]{.5\linewidth}
      \includegraphics[scale=0.42]{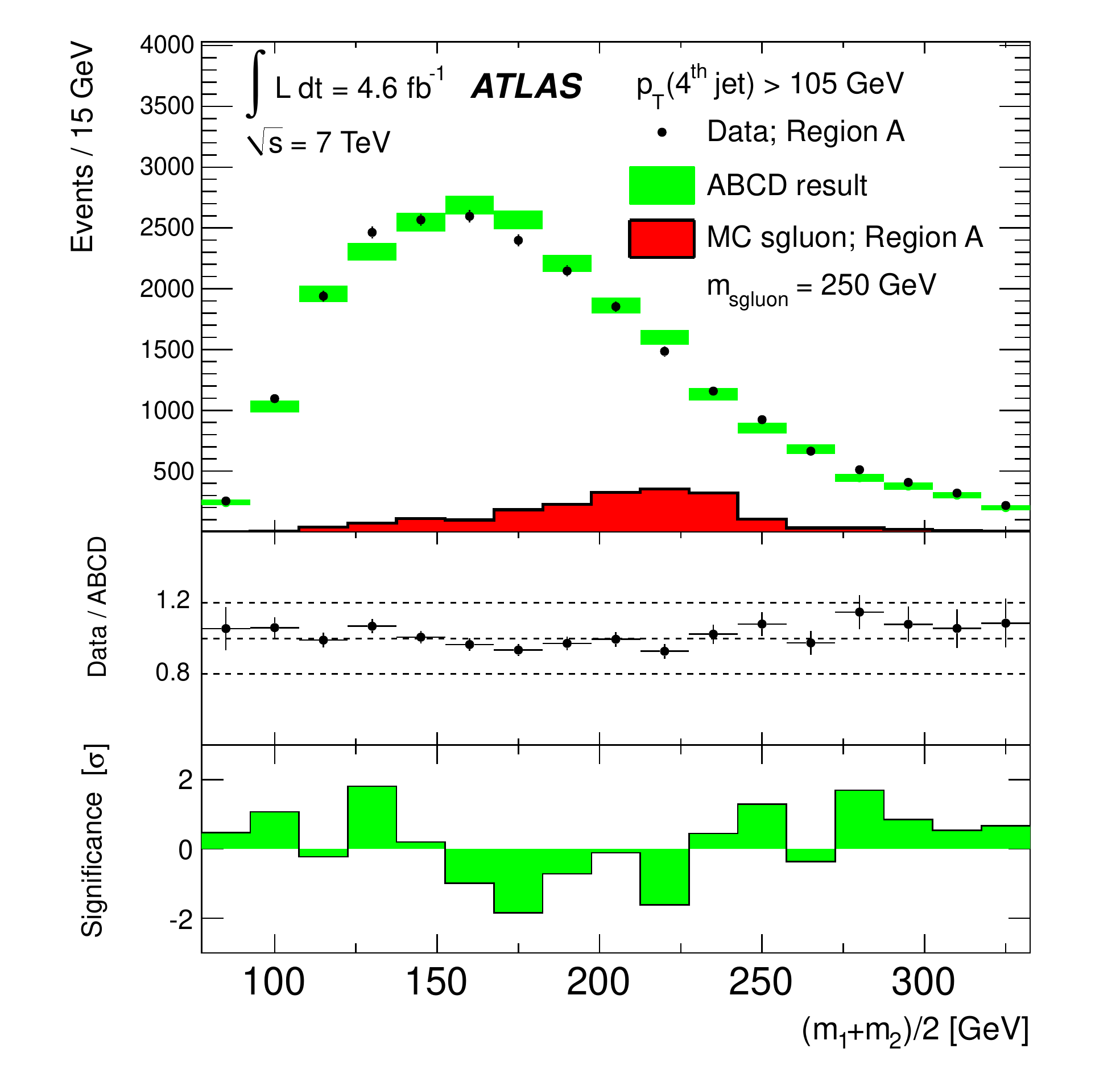}
   \end{minipage}\hfill
   \begin{minipage}[c]{\linewidth}
   \begin{picture} (12.,10.)(0.,0.)
		\setlength{\unitlength}{1.0cm}
		\put (4.5, 0.25){(a)}
		\put (13.5, 0.25){(b)}
   \end{picture}
   \end{minipage}
   \begin{minipage}[c]{.5\linewidth}
      \includegraphics[scale=0.42]{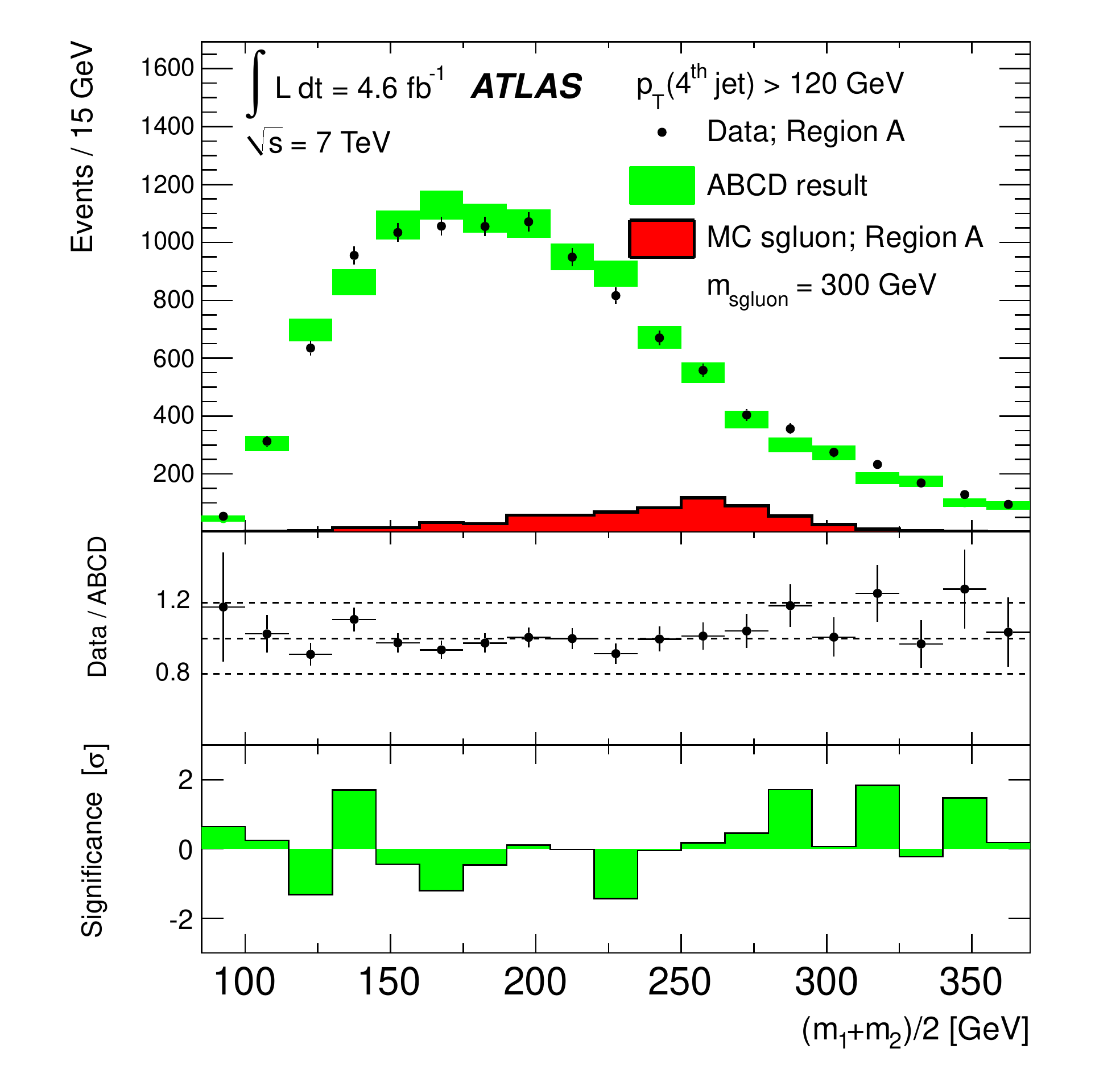}
   \end{minipage}\hfill
   \begin{minipage}[c]{.5\linewidth}
      \includegraphics[scale=0.42]{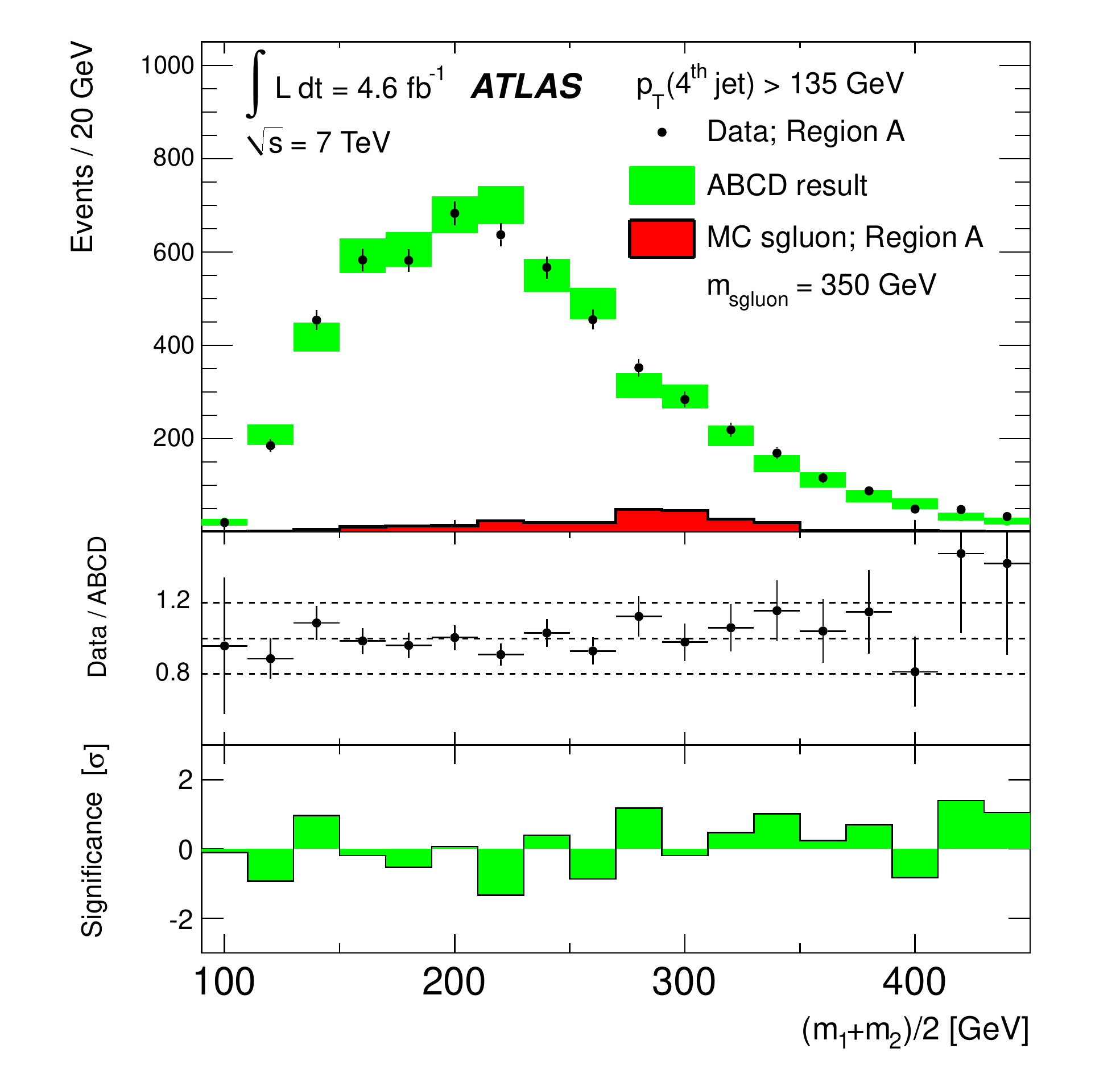}
   \end{minipage}\hfill
   \begin{minipage}[c]{\linewidth}
   \begin{picture} (12.,10.)(0.,0.)
		\setlength{\unitlength}{1.0cm}
		\put (4.5, 0.25){(c)}
		\put (13.5, 0.25){(d)}
   \end{picture}
   \end{minipage}
  \caption{
The comparison of the data in the signal region with the background prediction is shown for: 
(a) $m_{\mathrm{sgluon}}=150$\,\GeV, (b) $m_{\mathrm{sgluon}}=250$\,\GeV, 
(c) $m_{\mathrm{sgluon}}=300$\,\GeV\ and (d) $m_{\mathrm{sgluon}}=350$\,\GeV. 
The points are the data in the signal region (region A). The plain histogram (red) is the expected 
signal in region A normalised to the NLO cross-section. The prediction of background in region A 
based upon the data in region B normalised using the ABCD method is shown as the rectangles which include the statistical uncertainty. 
The data/background ratio and the statistical significance of its difference from one, in standard deviations, are also shown in the lower panels.}
  \label{fig:h0}
\end{figure*}
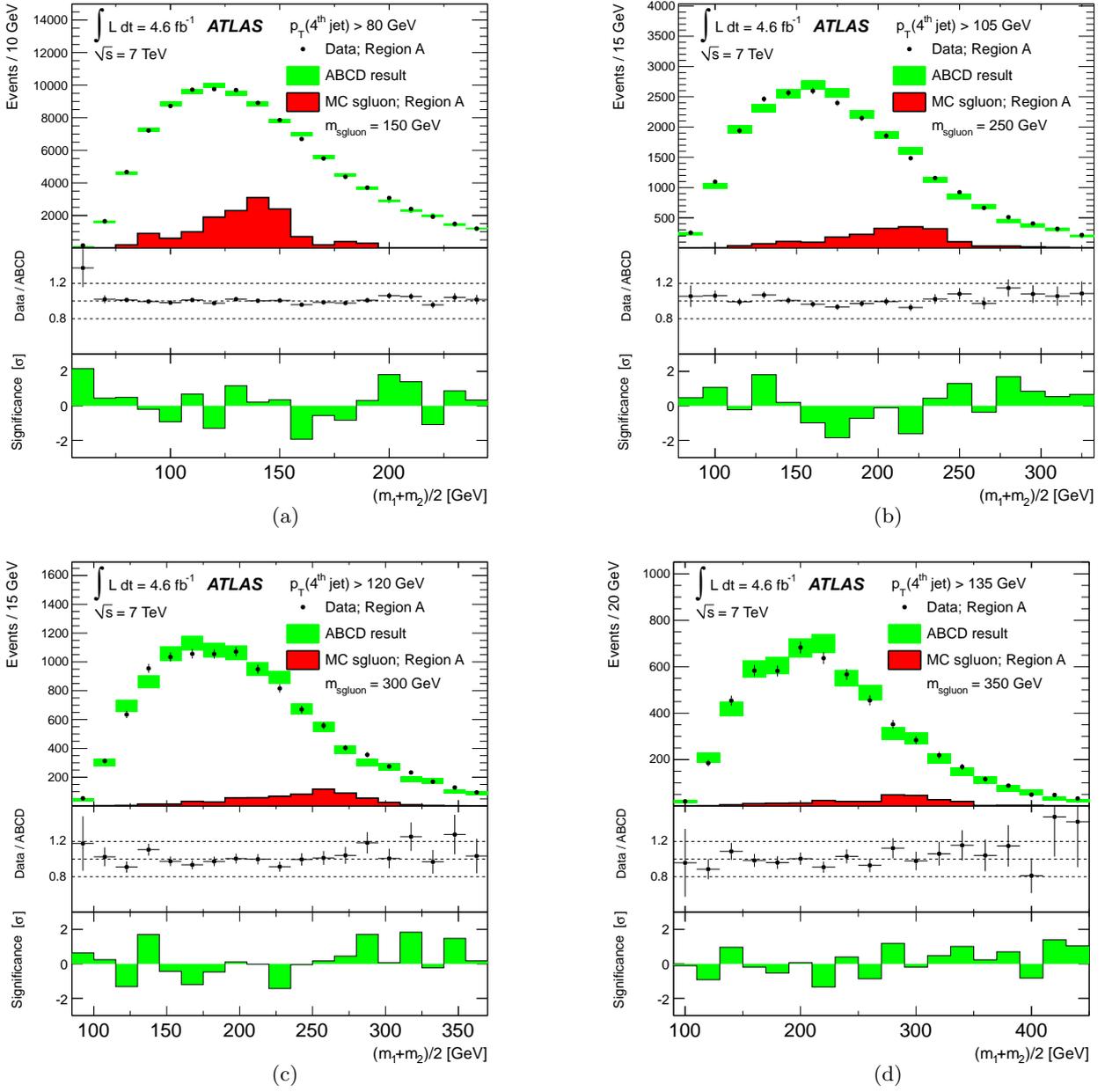

\begin{table*}[htb]
\caption{Comparison of the data in the signal region with the background prediction. The first column is the sgluon mass hypothesis, the second column is the corresponding minimum \pt\ requirement on the four jets, followed by the number of observed data events in the signal region~A (third column) and the number of predicted events using the ABCD method (fourth column), where the first uncertainty given is statistical and the second systematic. The last column gives the \textit{p}-value obtained from a Kolmogorov-Smirnov test between the shapes of the reconstructed average mass distributions in regions~A and~B. Only statistical uncertainties are considered in this test.}
\centering
\begin{tabular}{crrrc}
\hline
\hline
Sgluon Mass [GeV] & $p_\mathrm{T}^{\rm min}$ [GeV] & Data & ABCD prediction & Shape \textit{p}-value(A,B)  \\
\hline
 150 & 80     &    $102162$  &    $ 101100 \pm 800 \pm 2000$  &   0.22  \\
 200 & 90     &    $55194$   &    $ 54500  \pm 600 \pm 1100$  &   0.10  \\
 250 & 105    &    $23404$   &    $ 22500  \pm 340 \pm 500$   &   0.28  \\
 300 & 120    &    $11082$   &    $ 10640  \pm 230 \pm 210$   &   0.24  \\
 350 & 135    &    $5571$    &    $ 5330   \pm 180 \pm 110$   &   0.70  \\
\hline
\hline
\end{tabular}
\label{tab:matres}
\end{table*}

The correlation between the two variables is less than 0.1\% in the four regions in the data and less than 1\% in the \Pythia\ samples, so the normalisation of the background in the region~A is derived from the ratio of events in the control samples using 
$N_{\mathrm{A}}^{\prime}=N_{\mathrm{B}}\cdot N_{\mathrm{C}}/N_{\mathrm{D}}$.
A closure test is performed with the \Pythia\ and \Alpgen\ MC samples and shows that $N_\mathrm{A}^{\prime}$ reproduces the actual number of events in region A, $N_\mathrm{A}$, within 2\%. The difference is assigned as a systematic uncertainty on the background prediction.

Table~\ref{tab:matres} summarises the results obtained in data, for the five sgluon mass hypotheses and corresponding signal regions, A, together with the corresponding background predictions. Good agreement is observed. 
The assumption of the ABCD method that the shapes of the reconstructed average mass distributions in regions A and B are the same
was verified on the Monte Carlo samples. 
The last column gives the \mbox{\textit{p}-value} obtained from a Kolmogorov-Smirnov test 
between the shapes of the reconstructed average mass distributions for data in regions A and B.
Satisfactory \mbox{\textit{p}-values} are found
in this test on the data, which considers statistical uncertainties only.

The result of the background estimation is shown in Fig.~\ref{fig:h0} for sgluon masses of 150, 250, 300 and 350\,\GeV. 
The data in region A are compared to the data in the control region B normalised using the ABCD method. 
The expected sgluon signal in region A is also shown. 
The ratio of background to data as well as the significance, in standard deviations, of the difference between the data and the prediction
are shown as a function of the reconstructed average mass.
The significance takes into account only statistical uncertainties.

The average mass distribution of the data in region~A 
is compared to the background-only prediction and to the background-plus-signal prediction using a binned likelihood test, which incorporates the signal contamination in the control regions and systematic uncertainties via nuisance parameters. 
Systematic uncertainties affecting the simulated sgluon signal shapes are incorporated in the fit by varying signal templates,
taking into account the migration of events between the regions.

\begin{table*}[htb]
\caption{The systematic uncertainties on the signal due to the jet energy scale (JES), 
jet energy resolution (JER), initial and final state radiation (ISR/FSR), the trigger efficiency (Trigger), the Monte Carlo signal statistics (MC Statistics), the choice of parton distribution functions (PDFs) and  the integrated luminosity (Luminosity). 
The relative uncertainty of the signal acceptance is given for the four regions and for a sgluon mass of 300\,\GeV. 
The JES uncertainty is treated as asymmetric, corresponding to upward and downward fluctuations of the JES.
 For the JER uncertainty, only an upward fluctuation of the JER is considered,  i.e. only a degradation of the JER, with respect to the nominal MC JER. The last column shows the expected correlation among the four regions. 
}
\label{tab:sys}
\centering
\begin{tabular}{l||c|c|c|c|c}
\hline
\hline
Systematic & A & B & C & D & Correlation ABCD \\
\hline
JES       &  +10\%,--10\% &  +11\%,--11\%  &  +11\%,--13\% &  +15\%,--10\%  & 100\% \\
JER       &  +0\%,--2\%   &  +0\%,--7\%    &  +0\%,--1\%   &  +0\%,--2\%    & 100\% \\
ISR/FSR   &  +3.5\%,--3.5\%   &  +3.5\%,--3.5\%    &  +3.5\%,--3.5\%   &  +3.5\%,--3.5\%    & 100\% \\
Trigger       &  $\pm$1\%      &   $\pm$1\%      &  $\pm$1\%       &  $\pm$1\%      & 100\% \\
MC Statistics &  $\pm$4\%      &   $\pm$11\%     &  $\pm$5\%      &  $\pm$8\%       & 0\% \\
PDFs       &  $\pm$3\%      &   $\pm$3\%      &  $\pm$3\%       &  $\pm$2\%      & 0\% \\
Luminosity &  $\pm$3.9\%      &   $\pm$3.9\%      &  $\pm$3.9\%       &  $\pm$3.9\%      & 100\% \\
\hline
\hline
\end{tabular}
\end{table*}

The systematic uncertainties on the acceptance, and the correlation assumed for each uncertainty source between the four
regions, are listed in Table~\ref{tab:sys} for a sgluon mass of 300\,\GeV.
The uncertainty on the jet energy scale as well as the uncertainty on the jet energy resolution impacts both the signal shape and the acceptance.  
These uncertainties were measured with the complete 2010 dataset\\~\cite{atlas:jes} to which an extra uncertainty for the higher pile-up in the 2011 run was added.
The variation of the \Pythia\ parameters controlling initial state and final state radiation in a range consistent with experimental data~\cite{Aad:2012pw} produces only a small effect on the reconstructed average mass distribution. 
Therefore, this systematic uncertainty is taken into account for the signal acceptance. 
An uncertainty of 1\% is assigned to the trigger efficiency for sgluon signals.
The uncertainty on the signal acceptance due to the signal MC statistical uncertainty is uncorrelated among the four regions. 
The acceptance uncertainty due to the PDFs is estimated using the independent CT10~\cite{Lai:2010vv} error sets.
The uncertainty on the luminosity is taken to be 3.9\%~\cite{LUMINOSITYPLACEHOLDER,Aad:2011dr}.

To probe for the presence of a signal, a fit with a freely varying signal strength parameter is performed for each mass hypothesis. No significant deviation from zero is found, and limits on sgluon production are derived.

\begin{sloppypar} 
The profile likelihood ratio $\tilde q_\mu$~\cite{Cowan:2010js} is used as test statistic, and exclusions are determined using the CL$_s$ approach~\cite{Read:2002hq}.
Exclusion limits are computed using samples of \mbox{pseudo-experiments} generated taking into account all uncertainties and also contamination of the control regions by signal events. 
The normalisation and shape of the event distribution in region~B are used, whereas for regions C and D only the normalisation is used. 
MC templates are used to generate the shape of the signal in each pseudo-experiment. 
In each pseudo-experiment the statistical and systematic uncertainties are randomised, using Poisson and Gaussian distributions. 
\end{sloppypar} 

\begin{figure}[htbp]
	\center
	\includegraphics[width=\columnwidth]{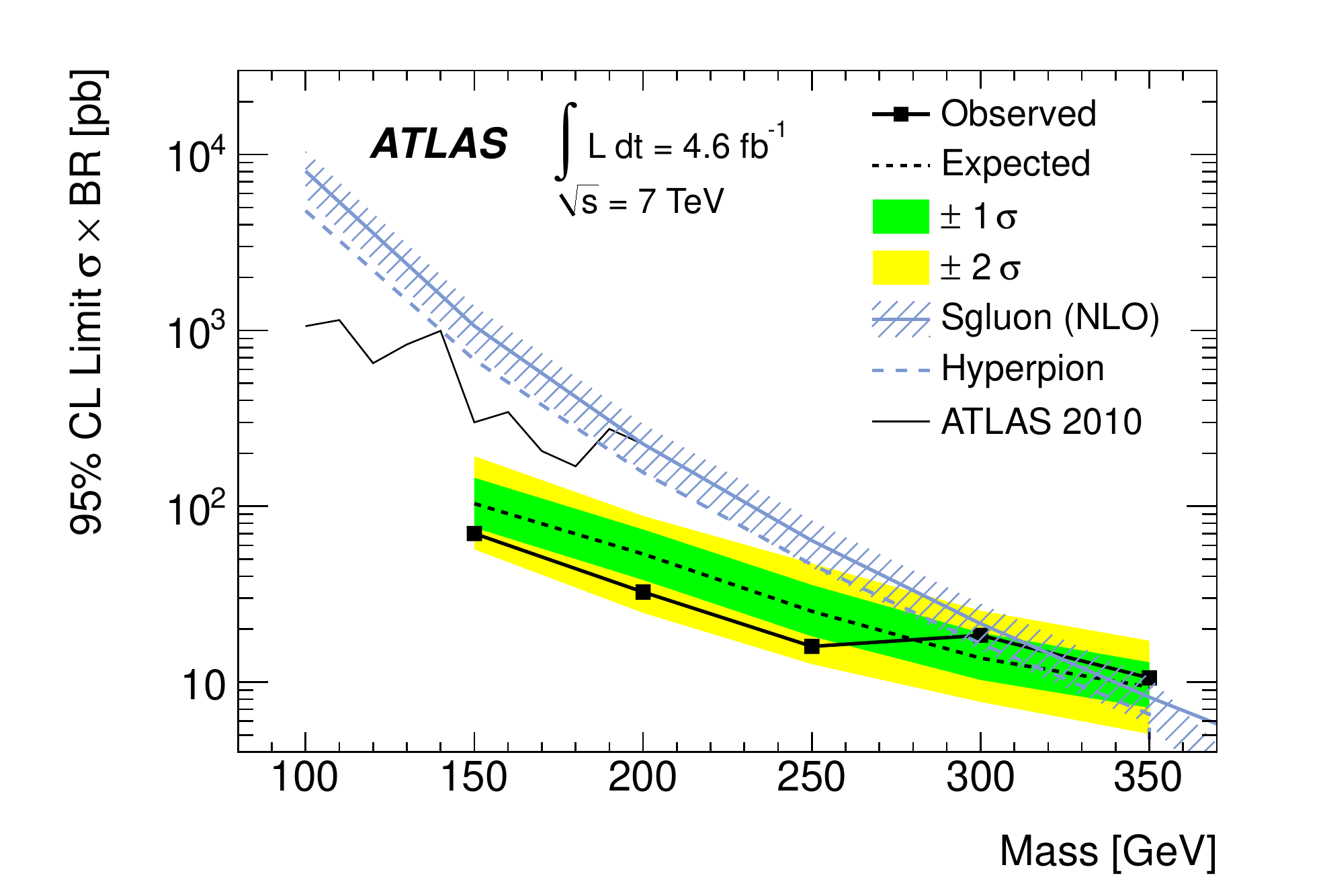}
	\caption{Expected and observed 95\% CL upper bounds on the product of the scalar pair production cross-section and 
the branching ratio to gluons as a function of the scalar mass using the profile likelihood ratio as the test statistic. 
The predictions of the sgluon and hyperpion pair production cross-section are also shown as well as the observed limit 
from a previous ATLAS search~\cite{Aad:2011yh}. The sgluon cross section is at NLO~\cite{GoncalvesNetto:2012nt}.}
	\label{fig:exclu_Xsec}
\end{figure}

Figure~\ref{fig:exclu_Xsec} shows the expected and observed 95\% CL upper bounds on the product of the scalar pair production cross-section and the branching ratio to gluons as a function of the scalar mass.
For a mass of 150\,\GeV\ (350\,\GeV), a limit of 70\,pb (10\,pb) on the scalar gluon pair production cross-section at the 95\% CL is obtained.
The solid line corresponds to the prediction of the sgluon pair production cross-section at NLO~\cite{GoncalvesNetto:2012nt}, 
which is larger than the leading order cross-section by a factor of about 1.6.
The hatched band indicates the systematic uncertainty due to the choices of renormalisation and factorisation scales.
Due to this recent NLO calculation, the previously unexcluded mass region around 140\,\GeV~\cite{Aad:2011yh} is now excluded by reinterpreting the limits obtained with the data recorded in 2010.
For the analysis of the data recorded in 2011, sgluons with a mass from 150\,\GeV\ to 287\,\GeV\ are excluded. The endpoint 
of the mass limit is defined as the intersection of the cross-section limit with the NLO cross-section minus one standard
deviation of the theory uncertainty.
 
The dashed line is the prediction for the hyperpion cross-section of a compositeness model, obtained by 
rescaling the sgluon cross-section according to the ratios from Ref.~\cite{Schumann:2011ji}. Since the ratios were calculated at leading order, this line should only be considered as an approximate indication of the excluded mass region.

\indent In summary, using 4.6\,\ifb\ of $\sqrt{s}=7$\,\TeV\ proton-proton collision data, collected by the ATLAS detector, four-jet events have been analysed to search for the pair production of a new scalar particle decaying into two gluons. 
The data in the signal region are in good agreement with the data-driven background estimation. 
No evidence for new phenomena is found. 
Cross-section times branching ratio limits as a function of the mass of the scalar particle are derived.
Interpreting the cross-section limit with the NLO cross-section calculation, sgluons with masses from 150\,\GeV\ to 287\,\GeV\ are excluded at 95\% CL using the data recorded in 2011 while the range 150\,\GeV\ to 306\,\GeV\ is expected to be excluded in the absence of a signal.
Combining the exclusions obtained with the analyses of 2010 and 2011 data, sgluons with masses from 100\,\GeV\ to 287\,\GeV\ are excluded at the 95\% CL.

We thank CERN for the very successful operation of the LHC, as well as the
support staff from our institutions without whom ATLAS could not be
operated efficiently.

We acknowledge the support of ANPCyT, Argentina; YerPhI, Armenia; ARC,
Australia; BMWF and FWF, Austria; ANAS, Azerbaijan; SSTC, Belarus; CNPq and FAPESP,
Brazil; NSERC, NRC and CFI, Canada; CERN; CONICYT, Chile; CAS, MOST and NSFC,
China; COLCIENCIAS, Colombia; MSMT CR, MPO CR and VSC CR, Czech Republic;
DNRF, DNSRC and Lundbeck Foundation, Denmark; EPLANET and ERC, European Union;
IN2P3-CNRS, CEA-DSM/IRFU, France; GNSF, Georgia; BMBF, DFG, HGF, MPG and AvH
Foundation, Germany; GSRT, Greece; ISF, MINERVA, GIF, DIP and Benoziyo Center,
Israel; INFN, Italy; MEXT and JSPS, Japan; CNRST, Morocco; FOM and NWO,
Netherlands; BRF and RCN, Norway; MNiSW, Poland; GRICES and FCT, Portugal; MERYS
(MECTS), Romania; MES of Russia and ROSATOM, Russian Federation; JINR; MSTD,
Serbia; MSSR, Slovakia; ARRS and MVZT, Slovenia; DST/NRF, South Africa;
MICINN, Spain; SRC and Wallenberg Foundation, Sweden; SER, SNSF and Cantons of
Bern and Geneva, Switzerland; NSC, Taiwan; TAEK, Turkey; STFC, the Royal
Society and Leverhulme Trust, United Kingdom; DOE and NSF, United States of
America.

The crucial computing support from all WLCG partners is acknowledged
gratefully, in particular from CERN and the ATLAS Tier-1 facilities at
TRIUMF (Canada), NDGF (Denmark, Norway, Sweden), CC-IN2P3 (France),
KIT/GridKA (Germany), INFN-CNAF (Italy), NL-T1 (Netherlands), PIC (Spain),
ASGC (Taiwan), RAL (UK) and BNL (USA) and in the Tier-2 facilities
worldwide.

\onecolumn
\clearpage
\begin{flushleft}
{\Large The ATLAS Collaboration}

\bigskip

G.~Aad$^{\rm 48}$,
T.~Abajyan$^{\rm 21}$,
B.~Abbott$^{\rm 111}$,
J.~Abdallah$^{\rm 12}$,
S.~Abdel~Khalek$^{\rm 115}$,
A.A.~Abdelalim$^{\rm 49}$,
O.~Abdinov$^{\rm 11}$,
R.~Aben$^{\rm 105}$,
B.~Abi$^{\rm 112}$,
M.~Abolins$^{\rm 88}$
% ...
\bigskip
\\
% ...
$^{11}$ Institute of Physics, Azerbaijan Academy of Sciences, Baku, Azerbaijan\\
$^{12}$ Institut de F{\'\i}sica d'Altes Energies and Departament de F{\'\i}sica de la Universitat Aut{\`o}noma de Barcelona and ICREA, Barcelona, Spain\\
% ...
$^{21}$ Physikalisches Institut, University of Bonn, Bonn, Germany\\
% ...
$^{48}$ Fakult{\"a}t f{\"u}r Mathematik und Physik, Albert-Ludwigs-Universit{\"a}t, Freiburg, Germany\\
$^{49}$ Section de Physique, Universit{\'e} de Gen{\`e}ve, Geneva, Switzerland\\
% ...
$^{88}$ Department of Physics and Astronomy, Michigan State University, East Lansing MI, United States of America\\
% ...
$^{105}$ Nikhef National Institute for Subatomic Physics and University of Amsterdam, Amsterdam, Netherlands\\
% ...
$^{111}$ Homer L. Dodge Department of Physics and Astronomy, University of Oklahoma, Norman OK, United States of America\\
$^{112}$ Department of Physics, Oklahoma State University, Stillwater OK, United States of America\\
% ...
$^{115}$ LAL, Universit{\'e} Paris-Sud and CNRS/IN2P3, Orsay, France\\
% ...
$^{s}$ Also at CPPM, Aix-Marseille Universit{\'e} and CNRS/IN2P3, Marseille, France\\
% ...
\end{flushleft}


\end{document}